\documentclass[%
 reprint,
 aps,
 pre,
 onecolumn,
longbibliography
]{revtex4-2}

\usepackage{graphicx}
\usepackage{dcolumn}
\usepackage{bm}
\usepackage{upgreek}
\usepackage{bm}
\usepackage{amssymb}
\usepackage{amsmath}

\begin{document}

\preprint{APS/123-QED}

\title{Coalescence of bubbles in a viscoelastic liquid}

\author{Alexandros T. Oratis}
\affiliation{Physics of Fluids Group, University of Twente, 7500 AE Enschede, The Netherlands}

\author{Vincent Bertin}
\affiliation{Physics of Fluids Group, University of Twente, 7500 AE Enschede, The Netherlands}

\author{Jacco H. Snoeijer}
\email[]{j.h.snoeijer@utwente.nl}
\affiliation{Physics of Fluids Group, University of Twente, 7500 AE Enschede, The Netherlands}

\date{\today}

\begin{abstract}
When two bubbles submerged in a liquid are brought closely together, the intermediate liquid film separating the bubbles begins to drain.
Once the film ruptures, the bubbles coalesce and form a neck that expands with time.
The dynamics of the neck growth are well understood in the context of pure, Newtonian liquids.
Yet, much less is known about the dynamics of this singularity when the surrounding liquid contains long flexible polymers, which provide viscoelastic characteristics to the liquid's properties.
Here, we experimentally study the coalescence of bubbles surrounded by polymer solutions.
In contrast to drop coalescence, and in spite of the singular stretching of polymers, we find that the presence of the dissolved polymers does not at all affect the coalescence dynamics at early times.
The polymer elasticity is found to slow down the flow only during the later stages of coalescence.
These observations are interpreted using an asymptotic solution of the Oldroyd-B model, which predicts a strong stress singularity near the extremity of the neck. However, the polymer stress turns out to diverge only in the azimuthal direction, which can explain why elastic effects remain subdominant during the initial stages of coalescence. \\
\end{abstract}

\maketitle


\section{Introduction}
Bubble coalescence is a physical process present in various industrial and environmental applications.
The ability of bubbles to coalesce as they collide in bubbly flows affects their size distribution, and can thus influence the production of cloud condensation nuclei \cite{neel2021collective}, volcanic eruptions \cite{gonnermann2007fluid,nguyen2013film}, or the properties of polymer foams \cite{taki2006bubble,andrieux2018liquid}.
From a fundamental point of view, bubble coalescence is of particular interest as an example of a hydrodynamic singularity: the change from two bubbles to a single bubble comes along with a change of topology of the fluid domain. 
Once two bubbles in close proximity establish contact, they form a microscopic neck.
In the immediate aftermath of coalescence, the curvature of the neck diverges.
The resulting singular capillary pressure leads to the retraction of the intermediate liquid film, which drives the growth of the neck as the two bubbles coalesce.
Depending on the liquid's properties, the growth of the neck is limited either by inertia or viscosity \cite{paulsen2014coalescence, munro2015thin, anthony2017scaling}.
The resulting inertio-capillary or visco-capillary balance gives rise to a rapid growth of neck, which increases with the square-root of time.
The associated prefactor of this scaling law depends on the relative importance of viscous to inertial effects, which can be characterized through the Ohnesorge number 
\cite{munro2015thin, anthony2017scaling, kamat2020bubble}. 
Different regimes of the spatial structure of the neck are identified, with intricacies depending on the fluid properties \cite{{munro2015thin, anthony2017scaling, kamat2020bubble}}.

In various natural and industrial settings featuring bubble coalescence, the surrounding liquid possesses viscoelastic properties \cite{gonnermann2015magma,ge2020analysis}.
It is thus of particular interest to better understand the manner in which viscoelasticity affects the dynamics during bubble coalescence. 
Indeed, viscoelastic liquids are ubiquitous in natural or industrial phenomena, and often contain long and flexible polymers.
When subjected to flow, the dissolved polymers get stretched and exert elastic stresses to the surrounding liquid.
After a critical inherent time scale, the relaxation time, the elastic stresses relax and the polymers return to their preferred coiled configuration.
Therefore, elastic stresses feature prominently in flows whose typical deformation rates occur much faster than the relaxation time~\cite{bird1987dynamics,larson2013constitutive,tanner2000engineering}. In the context of bubbles inside polymer solutions, viscoelastic stress gives rise to singular, cusp-like shapes at the rear of rising bubbles, changing dramatically the bubble's drag \cite{astarita1965motion,leal1971motion,soto2006study,fraggedakis2016velocity}.
A prime example  in the context of droplets is given by the  breakup of polymeric drops \cite{entov1997effect,anna2001elasto,clasen2006beads,eggers2020self,deblais2020self}. 
Rather than quickly pinching off, a very long thread forms that is sustained by highly stretched polymers. 
The thread thins exponentially in time, with a timescale dictated by the polymer relaxation times. 
Singularities, where stretching of fluid elements is very prominent, are therefore naturally of interest for viscoelastic liquids.

\begin{figure*}
\centering
\includegraphics[width=0.75\linewidth]{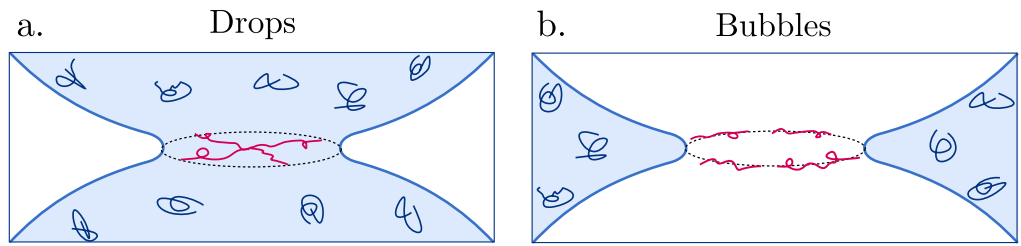}
\caption{ Cartoon of viscoelastic drop and bubble coalescence.
(a) When two drops coalesce, the expansion of the neck causes a dominant stretching in the radial direction near the neck.
(b) In contrast, the dominant stretching during bubble coalescence is in the azimuthal direction, leading to singular hoop stresses.
Note that the schematic displays dilute polymer solution, while the experiments are performed in the semi-dilute regime.
}
\label{fig:fig1}
\end{figure*}

In this paper we study the coalescence of bubbles immersed in viscoelastic fluids. 
This configuration is the natural counterpart of the coalescence of viscoelastic drops \cite{zdravkov2003film,yue2005diffuse}. 
The dynamics immediately after coalescence was studied in detail only recently \cite{dekker2022elasticity,bouillant2022rapid,varma2022elasticity}.
It was found that during the coalescence of semi-dilute polymeric drops,  polymer stretching only mildly influences the temporal evolution of the neck radius, as compared to Newtonian fluids. 
Yet, the dissolved polymers were shown to play an important role in the spatial shape adopted by the neck, due to the strong radial stretching of the polymers (see Fig.\,\ref{fig:fig1}a).
This led to the remarkable result that elastic stresses are highly singular, and locally dominant, but yet are globally negligible in how fast the neck grows. 
Here we focus on the analogous singularity of bubble coalescence in semi-dilute polymer solutions. 
Like drops, the dynamics of bubble coalescence involves a rapid growth of the neck resulting into singular polymer stretching. 
Interestingly, however, the flow structure differs significantly between the two cases, as bubble coalescence is driven by the retraction of the surrounding liquid film (see Fig.\,\ref{fig:fig1}b).
Specifically, we anticipate that polymer stretching during bubble coalescence will be primarily in the azimuthal direction rather than the radial. 

The paper is organized as follows. In Sec. II we first present the  experimental methods used to capture the coalescence between two bubbles.
We also report the rheology of the polyethylene-oxide (PEO) solutions used in this study, in terms of their relaxation time and viscosity.
In Sec. III we present our experimental results on the shape of the neck, in terms of its radius, self-similar features, and curvature.
Surprisingly, we find that the stretching of the polymers does not influence the neck dynamics at early times.
We rationalize the experimental observations in Sec. IV by deriving a force balance on the retracting viscoelastic film, including the elastic hoop force induced by the polymer stretching.
We find that the hoop stress becomes singular at the extremity of the neck when described by the elastic limit of the Oldroyd-B and model.
Yet, the singularity arises only in the azimuthal ``hoop'' direction and we show the stress singularity is integrable.


\section{Experimental methods}
\label{sec:Experiments}
To evaluate the effects of viscoelasticity on bubble coalescence, we perform experiments by recording the coalescence of two bubbles in polymer solutions using high-speed photography.
We first form a sessile bubble by slowly injecting air through a needle and bring it near a submerged hydrophilic surface.
A second bubble of the same radius is then injected through the needle and slowly brought into contact with the sessile bubble.
We vary the radii of the bubbles in the range $0.5<a<1.1$ mm, keeping the two bubbles at roughly the same initial radius.
When the bubbles are brought into close proximity, they appear to be in contact.
However, there still is an extremely thin intermediate film that separates the two bubbles and resists their merging~\cite{chan2011film}.
Once the film fully drains and ruptures, we record the coalescence from the side using a high-speed camera (Photron Nova S16) with frame rates reaching 150,000 frames per second.
The Navitar zoom lens mounted on the camera provides us a spatial resolution of approximately 3.5$\mu$m/pixel.
The interface profile of the neck is then extracted by custom image processing techniques, which we use to measure the evolution of the neck radius and shape.

To prepare viscoelastic liquids, we dissolve high-molecular weight polyethylene oxide (PEO, $M_w = 4\times10^6$ g/mol) in water. 
The solutions are mixed using a magnetic stirrer for at least 24 hours. 
The influence of polymers on the bubble coalescence is probed by varying the polymer concentration in the range $0.1<c<1$ wt.\%. 
Note that the lower concentration approximately coincides with the critical concentration where the polymer coils overlap. 
Varying the polymer concentration leads to significant changes of the rheological properties of the solution, such as the viscosity and relaxation time.

To measure the viscosity of each solution we use a rheometer (MCR 502 with CP50-1, Anton Paar) in a cone-plate configuration.
At low concentrations, the polymer solutions exhibit a fairly constant shear viscosity as a function of the imposed shear rate.
As we approach the critical overlap concentration $c \approx 0.1$ wt.\%, the polymer solutions adopt a shear-thinning behavior, as observed in Appendix~\ref{app:rheology}.
These solutions obtain a constant viscosity at low shear rates, which starts to decrease beyond a critical shear rate.
Specifically, the the zero-shear-rate viscosity of the solutions with concentrations $\{$0.1, 0.5, 1$\}$ wt.\% is measured to be $\{$4, 20, 340$\}$ mPa$\cdot$s respectively.

The relaxation time of the solution is measured by performing independent pinch-off experiments~\cite{anna2001elasto}.
While Newtonian drops pinch-off within a few milliseconds, the breakup of polymeric drops is largely delayed.
The presence of polymers leads to the formation of long liquid threads, which are stabilized by elastic stresses arising from the axial stretching of the polymers during the break-up.
Meanwhile, the elastic stresses relax and the threads thin under the action of capillarity.
Using the Oldroyd-B model to describe the rheology, the thread diameter is predicted to decay exponentially, with a characteristic decay time of $3\lambda$~\cite{bousfield1986nonlinear,bazilevskii1997failure,entov1997effect,chang1999iterated,anna2001elasto,clasen2006beads}.
We thus measure in-situ the relaxation time by following the temporal evolution of the thread's radius and fitting the decay rate (see Appendix \ref{app:rheology}).
The relaxation time of the solutions at concentrations $\{0.1$, $0.5$, $1\}$ wt.\% are found to be respectively $\{8, 20, 30\} \pm 3$ ms. 

The ratio of the relaxation time to the characteristic flow time scale is called the Deborah number and quantifies viscoelastic effects.
In bubble coalescence, the typical flow time scale is the inertio-capillary time scale $\tau = \sqrt{\rho a^3/\gamma}$, which depends on the liquid's density $\rho$ and surface tension $\gamma$.
The Deborah number $\lambda/\tau$ then ranges from 1 to 10.
However, the characteristic elongational rate at the early stages of bubble coalescence is much more violent, and can be estimated from the dynamics of the neck radius $r_0(t)$, according to $\dot r_0/r_0 \sim 1/t$ (see Section~\ref{sec:results}A).
It is thus more relevant to define an instantaneous Deborah number as $\lambda/t$~\cite{dekker2022elasticity}, which ranges from 10 to 1,000, where $t$ is set by the typical frame rate $\sim 10$ $\mu$s. Based on these large values we anticipate strong polymer stretching during coalescence. 


\section{Results}
\label{sec:results}
\begin{figure*}
\centering
\includegraphics[width=0.75\linewidth]{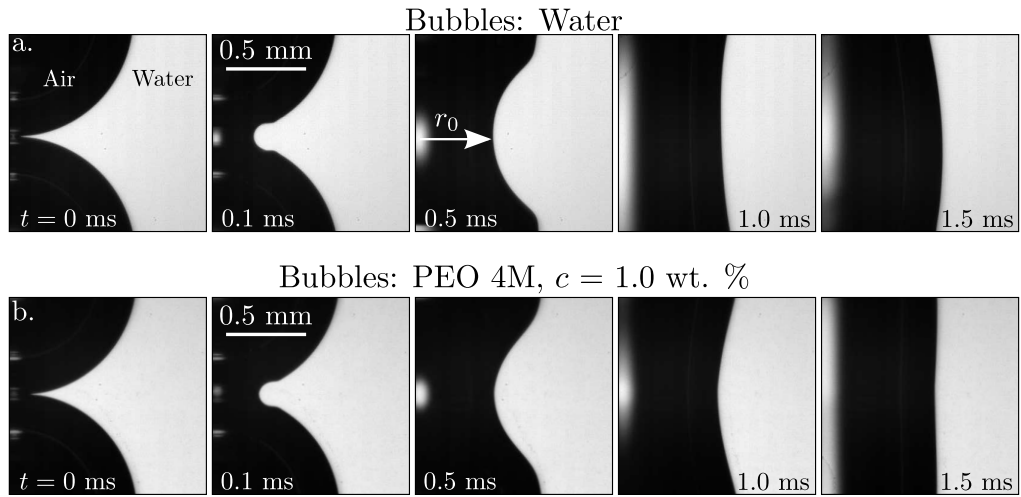}
\caption{Growth of the neck during bubble coalescence.
(a) When two bubbles coalesce in water, the liquid film retracts within milliseconds.
(b) Two bubbles in a polymer solution coalesce within a similar time scale. During the early stages of coalescence the neck shape and radius $r_0$ remain unaffected by the presence of polymers.
At the later times of coalescence the curvature of the neck appears to relax at a slower rate. This movie can be found in the Supplemental Material~\cite{Supp}, as well as a wide-view recording of the same process. }
\label{fig:fig2}
\end{figure*}

The process of bubble coalescence is illustrated in Fig.\,\ref{fig:fig2}, corresponding to Supp. Video 1~\cite{Supp}.
As the two bubbles establish contact, the thin film separating the bubbles begins to retract and the two bubbles coalesce to form a neck with radius $r_0$.
When the surrounding liquid is water, the bubbles coalesce within milliseconds.
At very early times, the rounded shape of the neck increases rapidly with time  (see Fig.\,\ref{fig:fig2}a).
At later times when the bubbles have merged sufficiently, the neck's shape changes from concave (positive curvature) to convex (negative curvature).
The situation is not very different for polymer solutions: even for the relatively high concentration of $c = 1$ wt.\%, the neck radius at early times seems to grow at a similar rate, and the neck shape appears to be almost identical (see Fig.\,\ref{fig:fig2}b).
We thus qualitatively observe similar coalescence dynamics between Newtonian and viscoelastic liquids at early times ($t \sim 0.1$ ms).
Surprisingly, this observation largely contrasts the results observed for viscoelastic drop coalescence, where the spatial structure of the neck was found to be substantially altered by the polymers \cite{dekker2022elasticity,bouillant2022rapid} (see also section~\ref{sec:drop}). For bubbles, only at later times do the polymers affects the coalescence dynamics where the change in curvature from concave to convex is delayed for the polymer solution ($t\sim 1$ ms).

\subsection{Neck radius}

\begin{figure*}[t]
\centering
\includegraphics[width=0.92\columnwidth]{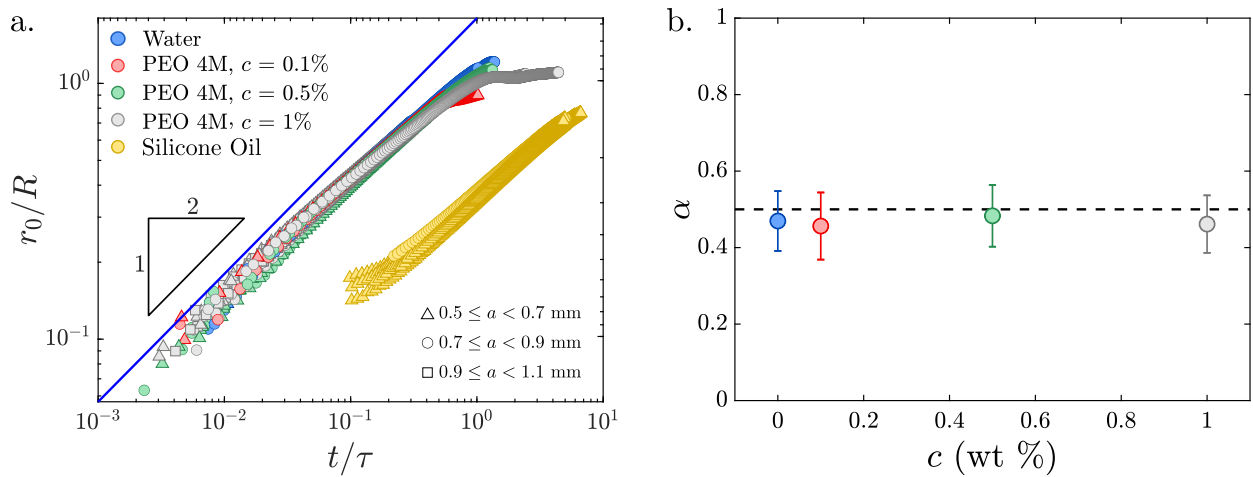}
\caption{ Temporal evolution of the neck radius.
(a) Plotting the neck radius $r_0$ normalized by the initial bubble radius $a$ against the dimensionless time $t/\tau$, we observe the data (circles) for water and polymeric solutions to fall on top of each other.
The data illustrates that the neck radius follows a power-law $r_0/a \sim (t/\tau)^\alpha$, with an exponent of $\alpha = 1/2$. The prefactor appears to be lower than suggested by theory for water [Eq.~\eqref{eq:keller}, solid line].
Conversely, the prefactor for bubbles coalescing in 350 mPa$\cdot$s silicone oil is much lower.
(b) Fitted values for the exponent $\alpha$. The exponent remains constant around the Newtonian value $\alpha = 0.5$ regardless of the polymer concentration $c$.
}
\label{fig:fig3}
\end{figure*}
We now turn to a quantitative analysis of the coalescence dynamics.
The growth of the neck radius $r_0$ normalized by bubble radius $a$ against dimensionless time $t/\tau$ is plotted in Fig.\,\ref{fig:fig3}a. 
The data for each polymer concentration and initial bubble radius falls on top of that of water. 
At early times $t/\tau< 0.2$, each data set is consistent with the scaling law $r_0\sim t^{1/2}$. This behavior is quantified by fitting curves of the form $r/a \sim (t/\tau)^{\alpha}$ on our experimental data for $t/\tau<0.2$, from which we extract exact exponent $\alpha$ in the immediate aftermath of the coalescence singularity.
As we increase the polymer concentration, the experimentally measured exponents remain fairly constant around $\alpha = 1/2$ (dotted line in Fig.\,\ref{fig:fig3}b).

It is of interest to compare these findings with their Newtonian counterparts \cite{paulsen2014coalescence,keller1983breaking,munro2015thin,anthony2017scaling}.
For Newtonian liquids, the prefactor of the neck radius scaling law is predicted to depend only on the Ohnesorge number $\mathrm{Oh} = \eta/\sqrt{\rho a\gamma}$, that corresponds to the ratio between the visco-capillary time $\tau_\mathrm{vis} = \eta a / \gamma$ and inertio-capillary time. 
The typical $\mathrm{Oh}$ of millimetric bubbles in water is on the order of $10^{-3}$, which corresponds to the limiting case where inertia dominates over viscosity. 
In this limit, the neck radius is predicted to follow \cite{paulsen2014coalescence,keller1983breaking,munro2015thin,anthony2017scaling}

\begin{equation}
	\label{eq:keller}
	r_0  = \left(\frac{32\gamma a }{3\rho} \right)^{1/4} t^{1/2} \approx 1.807  a \left(\frac{t}{\tau}\right)^{1/2}.
\end{equation}
We note that the experimental prefactor of the power law does not exactly match the inertial theoretical expectation (solid line in Fig.~\ref{fig:fig3}a). 
The measured prefactor is approximately $1.4$, which is similar to the value reported by previous experimental studies on bubble coalescence in low-viscosity fluids~\cite{thoroddsen2005coalescence,paulsen2014coalescence,soto2018coalescence}. 
The mismatch between thin-film theory and experiments for the prefactor has been attributed to finite-size effects~\cite{anthony2017scaling}. 
Indeed, high-speed photography experiments provide a spatial resolution about $\sim \mu\mathrm{m}$ and temporal resolution down to $10 \,\mu\mathrm{s}$, such that the neck radius at the first detected instant is already about $10\%$ of the initial bubble size, while the theory supposes a slender film~\cite{anthony2017scaling}.

The data in Fig.~\ref{fig:fig3}a for the polymer solutions are on top of that for water.
The zero-shear-rate viscosity of the three solutions used here are respectively 4, 20, and 340 times larger than water.
To test the effects of the liquid's viscosity, we also recorded the coalescence of bubbles surrounded by silicone oil with viscosity 350 mPa$\cdot$s, matching approximately the zero-shear-rate viscosity of 1 wt.\% PEO solution (see Fig.~\ref{fig:fig3}a.). 
The neck growth in viscous liquids still follows a $t^{1/2}$ scaling law~\cite{paulsen2014coalescence,munro2015thin}, but is significantly slower as compared to viscoelastic solutions (yellow data in Fig.~\ref{fig:fig3}a).
Therefore, the coalescence dynamics of these polymer solutions is not at all in a ``viscous" regime. 
This can be rationalised from the very high elongation rates of the flow during the film retraction. We estimate this to be of the order of $\dot{r}_0/r_0 \approx 1/(2t)\sim 1,000-10,000$ $\mathrm{s}^{-1}$, at which the viscosity of the solutions decreases dramatically (see Appendix~\ref{app:rheology}).
As a result, the effective viscosity of the solution is close to the solvent viscosity during the early times of coalescence.

Finally, at times beyond $t/\tau>0.2$ in Fig.~\ref{fig:fig3}, the growth of the neck starts to slow down and deviate strongly from the power law growth. 
During this interval, the neck has grown by a sufficient amount such that it almost reaches the initial bubble size. 
At this point, the azimuthal curvature becomes comparable to the tip curvature such that the flow field has a very different structure from the one described by the thin-film theory. 
The bubble now oscillates at its natural frequency $\sim 1/\tau$ and relaxes toward its spherical equilibrium shape. 
We notice that around this time, the bottom bubble detaches from the holding needle~\cite{soto2018coalescence} (see Supp. Video 2~\cite{Supp}). 
We can appreciate that the bubble pinch-off is significantly different from polymer solution to water, where long air cavities forms due to the elastic stresses~\cite{rajesh2022pinch}. 


\subsection{Neck shape}

\begin{figure*}
\centering
\includegraphics[width=0.99\columnwidth]{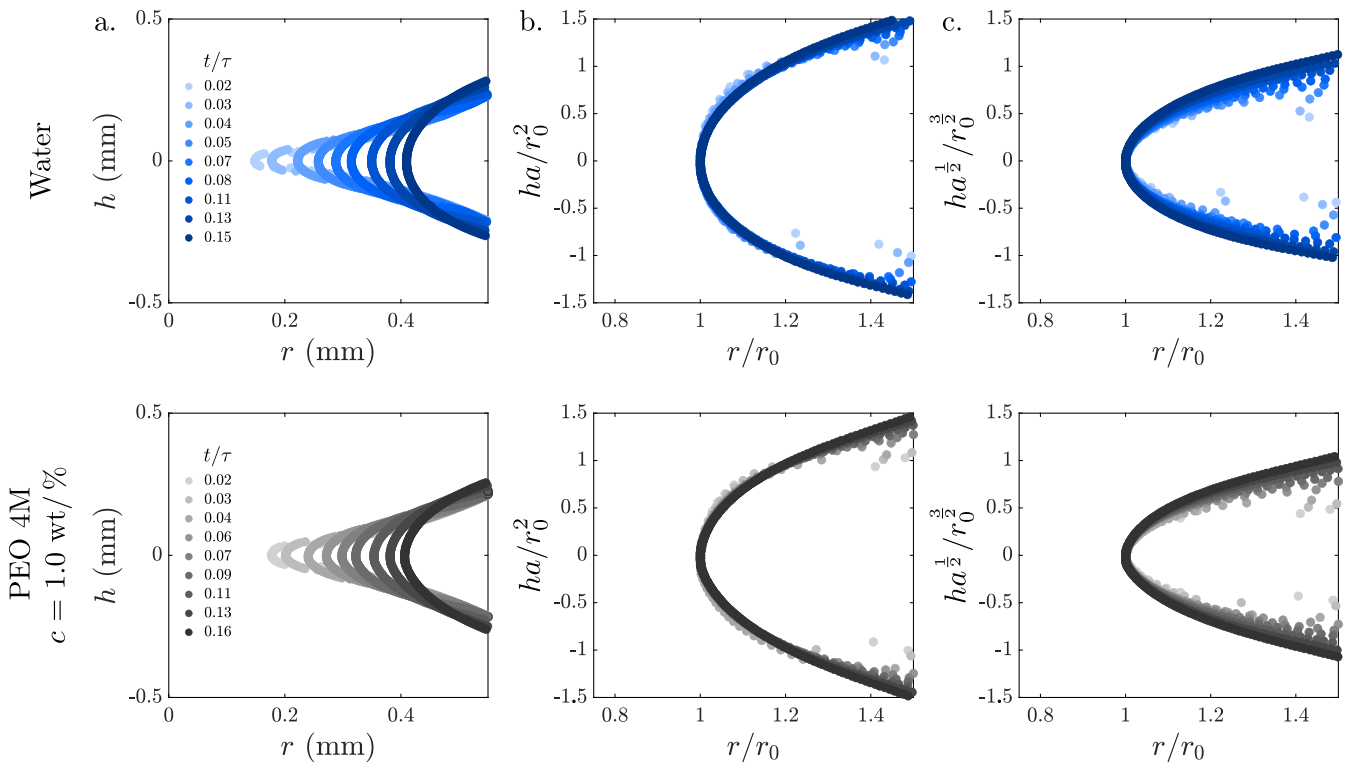}
\caption{Similarity profiles during bubble coalescence for water (top row) and polymer solution (bottom row).
(a) The dimensional profiles of the neck $h(r,t)$ against $r$ for temporal range of $0.02<t/\tau<0.16$.
(b) Using the parabolic scaling for the thickness $ha/r_0^2$ against the normalized radius $r/r_0$, the profiles collapse onto a single self-similar curve near the neck.
(c) The scaling for the thickness $ha^{1/2}/r_0^{3/2}$ does not collapse the data. This prediction is associated to the formation of a toroidal blob at very early times \cite{kamat2020bubble}, which might be present in a range not accessible in our experiments.
}
\label{fig:fig4}
\end{figure*}

Even though the temporal growth of the neck radius remains unaffected by the addition of polymers, the results from viscoelastic drop coalescence suggest that the neck profile could differ from the expected Newtonian behavior \cite{dekker2022elasticity}.
The instantaneous shape adopted by the neck at different times is reported in Fig.\,\ref{fig:fig4}.
We choose the two extreme cases of water and high-concentration polymeric solution for dimensionless time intervals of $0.02<t/\tau<0.16$.
The evolution of each profile as time progresses does not differ significantly between the two cases, with slightly sharper profiles near the tip for the viscoelastic liquid.
The rounded neck for both water and PEO solution appears to retain a similar shape during the retraction.
Normalizing the vertical coordinate of the sheet by the parabolic profile of the film $r_0^2/a$, and the radial coordinate by the neck radius $r_0$, we observe that each neck profile at different times collapses near the tip (see Fig.\,\ref{fig:fig4}b).

The self-similar properties of the liquid-film thickness thus follows the initial parabolic shape of the bubbles near the contact region prior coalescence. 
Previous theoretical models predict that the tip of the retracting film can adopt different scales, as it is rounded by capillarity. 
In the low-Ohnesorge limit, the retraction of the tip is mediated by its inertia and models find a significant mass accumulation in a toroidal blob of size $(r_0^3/a)^{1/2}$~\cite{kamat2020bubble}, similar to the shape adopted by a retracting film of uniform thickness \cite{brenner1999bursting,savva2009viscous,gordillo2011asymptotic}. 
However, in our experiments we can find no evidence for the accumulation of liquid in such a blob. 
The normalization of the film thickness by $(r_0^3/a)^{1/2}$ fails to collapse the data, as shown in Fig.\,\ref{fig:fig4}c. 
We propose that a toroidal blob may occur at scales that are not accessible in our experiment. 
As stated previously, the neck radius at the first detected instant is already about 10\% of the bubble radius.
Despite the spatio-temporal resolution down to $\sim$ 1 $\mu$m at $\sim$10 $\mu$s intervals, the spherical geometry of the interfaces limits the spatial radial scale to $\sim\sqrt{a h}\sim$ 10 $\mu$m.
Therefore, the fine details of the neck shape predicted theoretically at the very early-times are likely not accessible by optical experiments. 

\begin{figure*}[t]
\centering
\includegraphics[width=0.92\columnwidth]{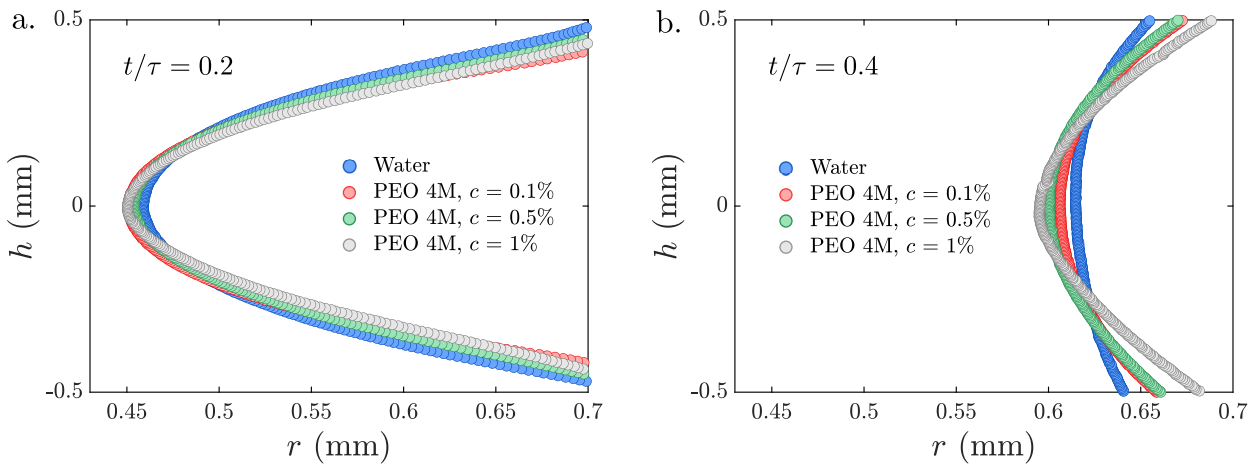}
\caption{(a) The dimensional profiles $h$ against $r$ at $t/\tau = 0.2$ for bubbles of radius $R = 0.8$ mm coalescing in solutions of various concentrations.
(b) As time progress to $t/\tau = 0.4$, the profiles start to differ and the curvature of the neck becomes sharper with an increase in the polymer concentration.}
\label{fig:fig5}
\end{figure*}

We now come to a direct comparison of the neck profiled for the different polymer concentrations. The dimensional profiles of the retracting shape for different polymer solutions is shown in Fig.\,\ref{fig:fig5}a. 
During the early stages of retraction at $t/\tau = 0.2$, we observe that all profiles nearly fall on top of each other. 
Therefore, we find experimentally that the polymers dissolved in the solution do not affect the shape during the early stages of coalescence. 
However, moments later, when $t/\tau = 0.4$, the shapes of the neck for the polymer solutions start to deviate from that of water (see Fig.\,\ref{fig:fig5}b). 
As the polymer concentration increases, the shape of the tip becomes sharper. 
We thus conclude that the polymer stretching plays a role in the coalescence dynamics after a certain time, or equivalently above a certain critical radius. 

\subsection{Neck curvature}

\begin{figure*}
\centering
\includegraphics[width=0.99\columnwidth]{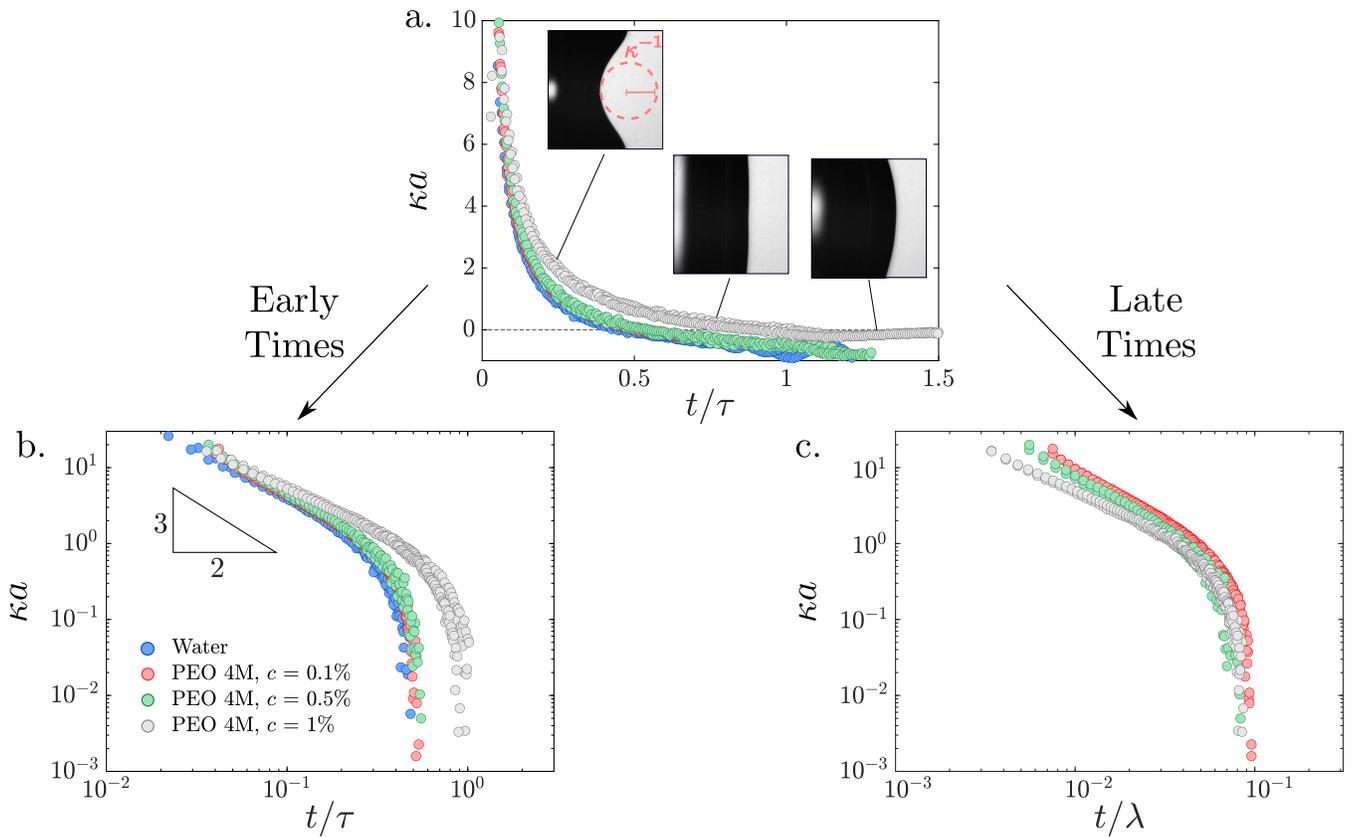}
\caption{The evolution of the tip curvature for coalescing bubbles with $a = 0.8$ mm.
(a)  Normalized neck curvature $\kappa R$ against dimensionless time $t/\tau$.
(b) Plotting these curves at logarithmic scales indicates that, at early times, the curvature collapses for water and all PEO solutions, and initially follows $\kappa \sim t^{-3/2}$.
(c) Normalizing time by the relaxation time collapses the data at late times.
Note that due to the sign change in curvature, the apparent curvature deceleration is a visual artefact when plotted in logarithmic scales.}
\label{fig:fig6}
\end{figure*}

To further quantify the effects of the dissolved polymers over the entire coalescence dynamics, we measure the tip curvature by fitting the proximity of the tip with circular functions. 
The normalized tip curvature against time is shown in Fig.\,\ref{fig:fig6}. 
The curvature initially decreases quickly before slowing down and switching from positive to negative once the neck radius has reached $a$. 
We can distinguish two dynamical regimes for the curvature: (i) the early-times dynamics driven by the retraction of the film, and (ii) a slow relaxation towards the final shape at late times. 
As discussed previously, in the early stages of coalescence $t/\tau <0.1$, the data for each polymer concentration falls on top of each other, when plotted versus a rescaled time using the inertio-capillary time. 
At these early times, the curvature decreases at early times as $\kappa\sim t^{-3/2}$ (see Fig.~\ref{fig:fig6}b).
The curvature of the tip can be expressed with the typical length scales of the interface profile such that $\kappa \sim r^*/h^{*2}$, where $r^*$ and $h^*$ are the radial and thickness scales of the tip respectively. 
From Fig.~\ref{fig:fig4}, we deduce that $r^*\sim r_0$ and $h^* \sim r_0^2$, leading to $\kappa \sim r_0^{-3} \sim t^{-3/2}$. 
Once again, the polymers initially have a minor influence on the evolution of the curvature. 
We stress that the apparent deceleration of the curvature dynamics observed with logarithmic scales does not correspond to any physical process. 
The curvature changes sign at a given time as the neck profile changes from concave to convex, which results in an apparent sharp decrease when the curvature is plotted on logarithmic scales.

In the late times of coalescence, for $t/\tau > 0.4$, the film has completely retracted and the bubble relaxes toward its equilibrium shape. 
In the experiments, the top bubble is attached to a hydrophilic surface, such that significant dissipation occurs at the contact line and the bubble oscillations are quickly damped, as compared to free bubbles~\cite{soto2018coalescence}.
Additionally, the bottom bubble systematically detached from the holding needle after a time typically of the order of $\tau$. Subsequent oscillations are generated by the bubble pinch-off (see Supp. Video 2~\cite{Supp}). 
Nevertheless, prior to the bubble pinch-off, we can observe in Fig.~\ref{fig:fig6}b that the tip curvature relaxation is delayed in polymer solutions, indicating elastic effects. 
Interestingly, the late-times curvature relaxation collapses when plotted versus a dimensionless polymeric time $t/\lambda$. 
A description of the corresponding dynamics is left for future work. 
We thus conclude that the characteristic time scale for describing the curvature dynamics is $\tau$ at early times and $\lambda$ at later times.
 
\begin{figure*}[t]
\centering
\includegraphics[width=0.92\columnwidth]{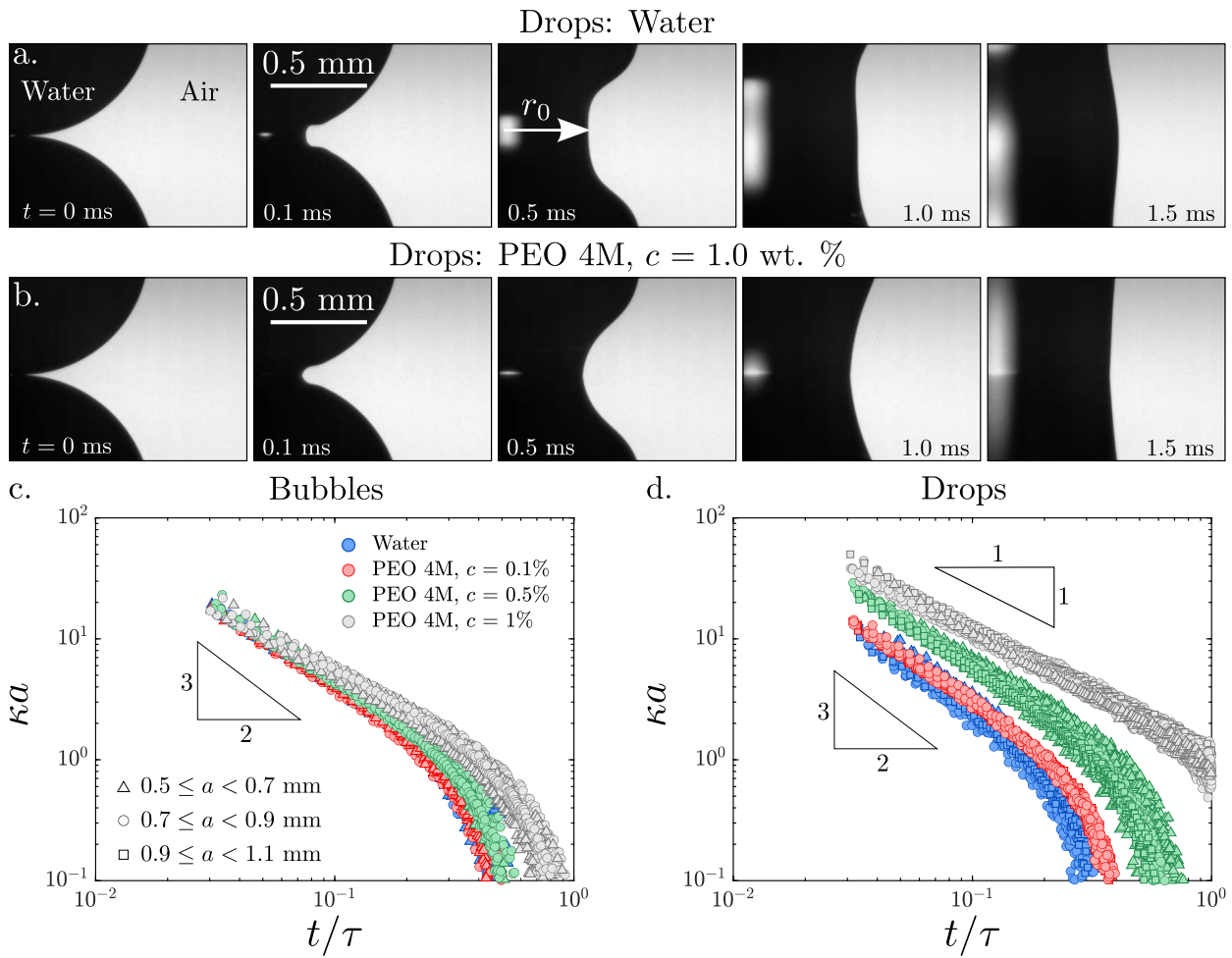}
\caption{Comparison between viscoelastic drop and bubble coalescence.
(a) When two water drops coalesce, the neck adopts a fairly flat shape that persists during the early stages of coalescence.
(b) The addition of PEO 4M with $c = 1$ wt.\% results in a much sharper shape of the neck during coalescence. 
This movie can be found in the Supplemental Material~\cite{Supp}.
(c) The addition of polymers slightly affects the curvature $\kappa$ of the neck during bubble coalescence.
(d) Conversely, the addition of polymers significantly affects the neck curvature during drop coalescence by orders of magnitude, already at early times.}
\label{fig:fig7}
\end{figure*}

\subsection{Comparison with drops}
\label{sec:drop}
The results of Sec.~\ref{sec:results} share some similarities with the dynamics of viscoelastic drop coalescence. 
The purpose of this section is to show a quantitative comparison between the two processes. 
We perform additional viscoelastic drop coalescence experiments, using a similar methodology as in~\cite{dekker2022elasticity}. 
Polymeric drops, using the same solutions as in Section~\ref{sec:Experiments}, were brought into contact and their subsequent coalescence were recorded with a high-speed camera.
In a similar manner as in Fig.~\ref{fig:fig3} for bubble coalescence, we recover that the addition of polymers has negligible effects on the growth dynamics of the minimum neck radius $r_0(t)$ (see Fig.\,\ref{fig:fig7}, corresponding to Supp. Video 3~\cite{Supp}). 
In contrast to Fig.~\ref{fig:fig5}, the shape of the neck during drop coalescence is significantly influenced by the addition of polymers. 
The neck adopts a fairly flat shape when two water drops coalesce (see Fig.~\ref{fig:fig7}a at $t = 0.5$ ms). 
Switching to a PEO solution with concentration $c = 1$ wt.\%, the neck shape is much sharper (see Fig.\,\ref{fig:fig7}b). 


To evaluate quantitatively the differences in the neck geometry between drop and bubble coalescence, we examine the early-times neck curvature, as in Fig.~\ref{fig:fig6}b.
For bubble coalescence, the differences between water and polymeric solutions are minimal and the curvature scales as $\kappa \sim t^{-3/2}$.
In contrast, the effects of the polymers on the curvature during drop coalescence are much more striking (see Fig.\,\ref{fig:fig7}d).
As the amount of dissolved polymer is increased, the values of the curvature increases up to an order of magnitude for 1 wt.\% concentration, reflecting the sharpening of the tip. 
Additionally the temporal decrease of the curvature is modified by the polymer from $t^{-3/2}$ to what appears to be a $t^{-1}$ regime, as the polymer concentration increases~\cite{bouillant2022rapid}. 

The differences between drop and bubble coalescence can be attributed to the very different flow fields.
In drop coalescence, the polymer stretching is in the radial direction (see Fig.~\ref{fig:fig1}a), leading to radial elastic stresses $\sigma_{rr}$. 
The latter balances the Laplace pressure at the tip interface $\gamma\kappa \sim \sigma_{rr}$, which locally modifies the shape of the neck. 
Assuming the polymer stresses to be caused by extensional viscosity as $\sigma_{rr} \sim \bar \eta_\infty \dot{r}_0/r_0 \sim \bar\eta_\infty / t$, where $\bar \eta_\infty$ is the extensional viscosity at high rate, we obtain a curvature temporal decay as $\kappa \sim t^{-1}$~\cite{bouillant2022rapid}. 
In bubble coalescence, the elastic stresses are primarily in the azimuthal direction, and generate elastic hoop forces. 
However, the normal stress balance at the bubble tip interface does not involve the azimuthal stress. 
This offers an explanation why, in contrast to drops, the neck shape in bubble coalescence is not altered by viscoelasticity at early times (see Fig.~\ref{fig:fig5}a). 
In the next section, we model the elastic hoop force occurring at the tip of the bubble and discuss its influence on the growth dynamics. 

\section{Bubble coalescence model}
\label{sec:theory}
\subsection{Tip force balance}

\begin{figure}[t!]
\centering
\includegraphics[width=0.75\linewidth]{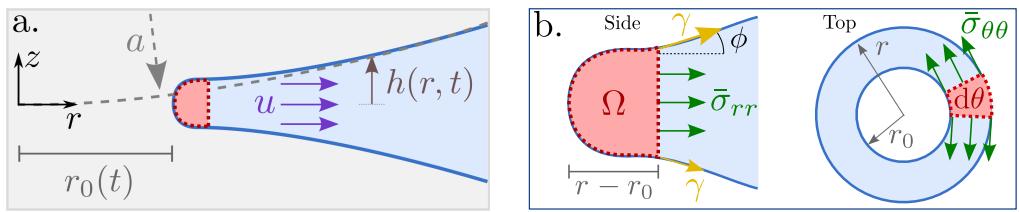}
\caption{
(a) When two bubbles with initial radius $a$ coalesce, they cause a radial flow concentrated near the tip with velocity $u$.
The retraction is characterized by the film of thickness $h(r,t)$ and the radial extend $r_0(t)$. 
(b) To model the retraction, we consider the forces acting on a control volume $\Omega$, of typical size $r - r_0$. 
These forces include the surface tension $\gamma$ as well as the radial and hoop stresses $\sigma_{rr}$ and $\sigma_{\theta\theta}$, exerted by the polymers.
}
\label{fig:fig8}
\end{figure}

We now model the elastic response of polymers during the early stages of bubble coalescence.
We follow the approach of previous studies~\cite{munro2015thin,kamat2020bubble}, which demonstrate that the coalescence dynamics can be understood by expressing a global force balance on the tip, matched to a slender description of the liquid film (Fig.~\ref{fig:fig8}). 
Our analysis focuses  on the tip region, where polymer stretching is largest and based on which the dynamics can be understood in the inertial regime \cite{keller1983breaking}. We introduce a control volume denoted $\Omega$, defined by an arc of angle $\mathrm{d}\theta$, from the tip position $r_0$ to an arbitrary radial position $r$ (see Fig.~\ref{fig:fig8}). 
Then, the radial force balance per azimuthal length takes the form

\begin{equation}
    \frac{dP}{dt} = \gamma r_\mathrm{E} \cos \phi +   \left.r h \bar{\sigma}_{rr}\right|_{r = r_\mathrm{E}} - \int_{r_0}^{r_\mathrm{E}}h\bar{\sigma}_{\theta\theta}\,dr,
    \label{eq:Force_Balance_exact}
\end{equation}
where we introduce the total radial momentum $P$ as 
\begin{equation}
 \frac{dP}{dt} = \int_\Omega \rho \frac{D u}{D t} \,\mathrm{d}^3x,
 \end{equation}
and the thickness-averaged radial and hoop fluid stresses $\bar{\sigma}_{rr}$ and $\bar{\sigma}_{\theta \theta}$ respectively. 
Here, the radial velocity is denoted $u$. 

Equation~\eqref{eq:Force_Balance_exact} is an exact integral of the radial momentum equation without approximation; yet, one needs to model the stresses and velocity field to capture the neck dynamics. 
In the low-Oh limit, the viscous stresses in the tip are negligible, leaving only the polymer contribution in the stress. 
For Newtonian liquids, where the elastic hoop force is absent, the liquid accumulates in an annular blob of size $\sim r_0^{3/2}/a^{1/2}$, as obtained by volume conservation.
Therefore, using the volume $\Omega\sim r_0^4/a$ for the retracting tip~\cite{keller1983breaking}, we recover Eq.~\eqref{eq:keller} for the temporal growth of the neck radius. 
Below we estimate the polymer stress and whether it effects the dynamics of the tip.


\subsection{Polymer stress}

In what follows, we assume, like in \cite{keller1983breaking}, that the radial velocity in the tip region is nearly uniform and given by $\mathbf{u} \approx u \mathbf e_r = \dot{r}_0 \mathbf{e}_r$. 
This velocity field exhibits a gradient $\mathbf{\nabla}\mathbf{u} \approx \frac{u}{r} \mathbf{e}_\theta \mathbf{e}_\theta$, such that the stretching of fluid elements occurs only in the azimuthal direction. 
By consequence, the polymers in the tip region will predominantly be stretched in the ``hoop" direction, as drawn in Fig.~\ref{fig:fig1}. 
The absence of stretching in the radial direction implies a vanishing radial elastic stress, $\bar\sigma_{rr} \approx 0$, which is consistent with the observation that the neck curvature is unaffected by polymers. 
Namely, the jump in normal stress $\gamma \kappa$ involves the radial component $\bar\sigma_{rr}$ and not $\bar\sigma_{\theta\theta}$. 
In the remainder we thus focus on $\bar\sigma_{\theta\theta}$, and compute its effect on the momentum equation \eqref{eq:Force_Balance_exact}.

\subsubsection{Elastic limit of Oldroyd-B}


To evaluate the contribution of the polymers during bubble coalescence we first consider the Oldroyd-B model, which has been successfully employed to analyze the stretching dynamics of thin viscoelastic liquids jets \cite{renardy1995numerical,entov1997effect,clasen2006beads,turkoz2018axisymmetric,deblais2020self,eggers2020self}, as well as their retraction~\cite{sen2021retraction}.
The polymeric stress tensor can be expressed as $\boldsymbol{\sigma}=G(\mathbf{A}-\mathbf{I})$, where $G$ is the polymer shear modulus and \textbf{A} is the conformation tensor that describes the amount of polymer deformation ~\cite{bird1987dynamics}. 
The formulation of the Oldroyd-B model is completed with a relaxation equation $\overset{\triangledown}{\mathbf{A}} = -(\mathbf A - \mathbf I)/\lambda$, where the upper convected derivative is defined as

\begin{equation}
    \label{eq:upperconvected}
   \overset{\triangledown}{\mathbf{A}} = \frac{\partial \mathbf{A}}{\partial t} + (\mathbf{u}\cdot \mathbf{\nabla})\mathbf{A} - (\mathbf{\nabla}\mathbf{u})^T\cdot \mathbf{A} - \mathbf{A} \cdot \mathbf{\nabla}\mathbf{u}.
\end{equation}
Importantly, we recall that near the tip $\mathbf{\nabla}\mathbf{u} \approx \frac{u}{r} \mathbf{e}_\theta \mathbf{e}_\theta$. 
This implies that the evolution equation for $\mathbf A$ is indeed only driven in the azimuthal direction, while radial stretching is absent. 

An exact solution for the polymer stretching can be obtained when using that the typical stretching rate $u/r = \dot r_0/r_0 \sim 1/t$, which at early times is much larger than the relaxation rate $1/\lambda$. 
In this limit, the polymers respond purely elastically, and their evolution can be inferred from $\overset{\triangledown}{\mathbf{A}}=0$ \cite{snoeijer2020relationship}. 
The resulting equation for the azimuthal component $A_{\theta\theta}$ is then given by 

 \begin{equation}
    \label{eq:Oldroyd_B_azimuthalbis}
    \frac{\partial A_{\theta\theta}}{\partial t} + u \frac{\partial A_{\theta\theta}}{\partial r}-\frac{2u}{r}A_{\theta\theta}  = 0.
\end{equation}
This equation can be solved analytically using the correspondence to elastic solids \cite{snoeijer2020relationship}. Using a stress-free initial condition, we find (see Appendix~\ref{app:Oldroyd}): 

\begin{equation}
    A_{\theta\theta}(r,t) = \frac{r^2}{\left[8a\int_{r_0(t)}^r \tilde{r}h(\tilde{r},t) d\tilde{r} \right]^{1/2}}.
    \label{eq:A_theta}
\end{equation}
One indeed verifies that this is a solution to \eqref{eq:Oldroyd_B_azimuthalbis} upon invoking mass conservation $r \dot{h} + \frac{\partial}{\partial r} (r h u) = 0$; the stress-free condition is verified for the initial shape $h(r,t=0) = r^2/(2a)$, for which $A_{\theta\theta}=1$. 
Note that within the context of the Oldroyd-B model, the above expression offers an upper bound to the polymer stress; the omitted relaxation will only reduce the stress. 

With the expression for the conformation tensor in hand, we can find the azimuthal polymer stress $\bar{\sigma}_{\theta\theta}$ in the tip. 
We first remark that setting $r=r_0$ in \eqref{eq:A_theta} gives a vanishing denominator: the stretching at the tip is predicted to be infinite at the neck radius for the Oldroyd-B model. 
We will show below that this singularity can be lifted when invoking nonlinear relaxation -- for now, we exploit that this singularity is integrable, so that the hoop force inside the tip remains finite even in the Oldroyd-B model. 
Hence, we proceed by estimating the stretch inside the tip region by assuming a typical tip size $h^*$, that is is smaller than the neck radius $r_0$. 
This approach gives $A_{\theta\theta}\sim r_0^{3/2}/(a^{1/2}h^*)$, so that the azimuthal polymer stress becomes
\begin{equation}
\bar{\sigma}_{\theta\theta}\sim G\left(\frac{r_0^{3/2}}{a^{1/2}h^*}-1\right) \sim \frac{G r_0^{3/2}}{a^{1/2}h^*}.
\end{equation} 
Assuming the tip thickness scales as $h^*\sim r_0^2/a$, as observed experimentally (see Fig.~\ref{fig:fig4}b), we find that the hoop stress takes the form $\bar{\sigma}_{\theta\theta} \sim G(a/r_0)^{1/2}$. 
Indeed, the typical azimuthal polymer stress inside the tip diverges at early times after coalescence, as $r_0 \rightarrow 0$. 

To estimate the effect of elasticity on the momentum balance Eq.~\eqref{eq:Force_Balance_exact}, we obtain a scaling expression for the polymeric hoop force

\begin{equation}
	\label{eq:hoop-force}
\int_{r_0}^{r} h\bar{\sigma}_{\theta\theta} \, \mathrm{d}r \sim \bar{\sigma}_{\theta\theta}h^{*2} \sim G\,\frac{r_0^{3/2}h^* }{a^{1/2}}.
\end{equation}
While the hoop stress diverges when $r_0 \rightarrow 0$, the hoop force vanishes. Yet, as the film begins to retract and both $r_0$ and $h^*$ grow with time, the hoop force can significantly increase. 
The remaining question is whether the elastic hoop force may overcome the capillary forces that drives the retraction. 
Comparing the hoop force \eqref{eq:hoop-force} to the capillary term in Eq.~\eqref{eq:Force_Balance_exact}, which scales as $\gamma r_0$, we find that the elastic hoop force would only become comparable to surface tension once the neck has retracted by a distance

\begin{equation}
    r_0 \sim \left(\frac{\gamma/G}{h^*} \right)^2a,
\end{equation}
which involves the elastocapillary length $\gamma/G$. 
For a typical shear modulus of $G = 10$ Pa found for PEO solutions~\cite{ebagninin2009rheological}, the elastocapillary length $\gamma/G$ is of the order of a millimeter, which is much larger than $h^*$ in our experiment.  
Hence, elastic hoop forces in the coalescence experiments are always much smaller than capillary forces. 

As mentioned previously, the denominator in  Eq.~\eqref{eq:A_theta} vanishes when setting $r = r_0$,  leading to a singular azimuthal stretch $A_{\theta\theta}$.
Consequently, the azimuthal stress $\bar{\sigma}_{\theta\theta} \approx GA_{\theta\theta}$ also diverges as we approach the extremity of the tip.
If one supposes that the extremity of the tip is circular and scales as $h(r) \propto (r-r_0)^{1/2}$, the hoop stress will then scale as $\bar{\sigma}_{\theta\theta} \propto (r-r_0)^{-3/4}$. 
This stress singularity is weak and integrable, such that the hoop force defined in Eq.~\eqref{eq:Force_Balance_exact} is well defined.
As a result, we can integrate the hoop stress over the entire tip surface to compute the hoop force, which scales as $\bar{\sigma}_{\theta\theta} h^{*2}$.
Therefore, while the elastic limit of the Oldroyd-B model predicts a singular hoop stress in the tip extremity, the resultant hoop force in the radial momentum equation is subdominant.


\subsubsection{Nonlinear relaxation}

It is obviously of interest to explore the robustness of the above statements for constitutive models beyond the Oldroyd-B fluid. 
Indeed, a feature of the Oldroyd-B model is that infinite stresses occur at large extensional rates.
Yet, experimental studies demonstrate that the extensional viscosity of polymer solutions saturates to a value $\bar\eta_\infty$ in this limit \cite{anna2001elasto}.
Therefore, we hypothesize that the stress at the extremity of the neck saturates to a value $\bar\sigma_{\theta\theta} \sim \bar \eta_{\infty} u/r \sim \bar \eta_\infty/t$, which resolves the blow-up of stress as $(r - r_0) \to 0$. 
This scenario can be obtained analytically when incorporating nonlinear relaxation (e.g. Giesekus model) or finite extensibility of the polymers (FENE-P model)~\cite{bird1987dynamics}.
Therefore, considering more complex constitutive laws would effectively lower the stress as compared to the Oldroyd-B model and not modify the conclusion of our analysis.

\section{Discussion}

To summarize, we have investigated the retraction of a low-viscosity viscoelastic film that occurs during the coalescence of two bubbles.
Using experiments and a theoretical model of the polymer stresses, we focused on the spatio-temporal evolution of the retracting film during the early phase of coalescence. 
The latter is found to adopt a self-similar shape at early times in the experimentally accessible range. 
Interestingly, the retraction velocity and the shape adopted by the film is not affected by the viscoelastic nature of the solution. 
Throughout the retraction, the polymers get stretched azimuthally and generate an elastic hoop force. 
We showed that while the hoop stress is highly singular, the hoop forces are asymptotically smaller than the capillary forces driving the retraction. This renders bubble coalescence fundamentally different from drop coalescence, for which the spatial structure of the neck is strongly altered by elasticity~\cite{dekker2022elasticity,bouillant2022rapid,varma2022elasticity}. 
For bubbles, we observe viscoelastic effects in the late stages of coalescence, where the relaxation of the bubble toward its equilibrium shape is delayed by viscoelasticity. 

\section*{Acknowledgements}
We acknowledge A. Bouillant, C. Datt, J. Eggers and V. Sanjay for stimulating discussions. This work is supported by the N.W.O through the VICI Grant No. 680-47-632.

\appendix
\section{Rheological measurements}
\label{app:rheology}
\begin{figure*}[t]
\centering
\includegraphics[width=0.92\columnwidth]{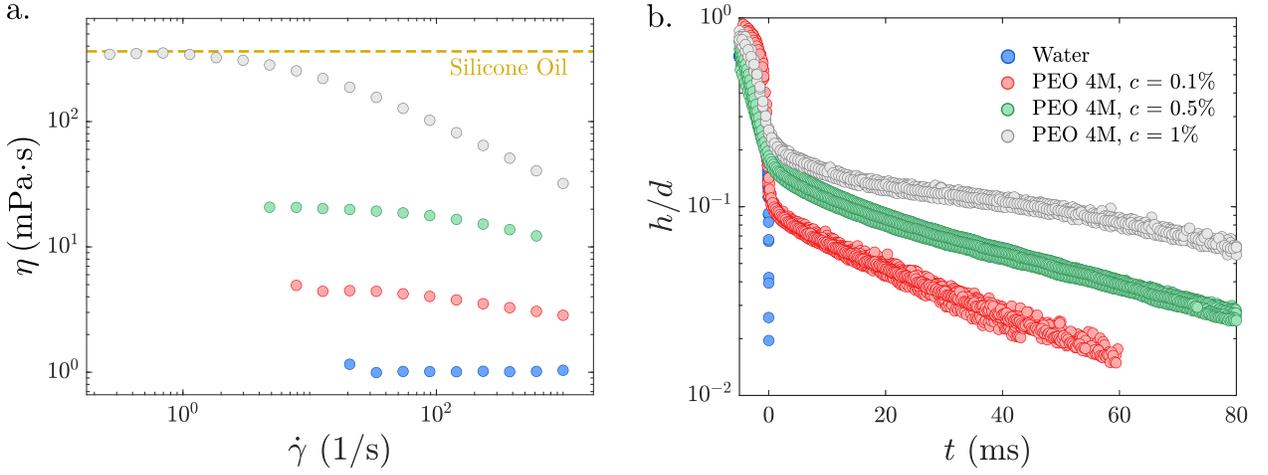}
\caption{Rheological characterization of polymer solutions.
(a) The measured viscosity $\eta$ plotted against the applied shear rate $\dot{\gamma}$.
The dotted line corresponds to the viscosity of the silicone oil viscosity used in Fig.\,\ref{fig:fig3}, as reported by the manufacturer.
(b) Semi-logarithmic plot of the thread thickness as a function of time during a drop pinch-off experiment. 
For polymer solutions, the thread thickness decays exponentially, with a characteristic decay time proportional to the relaxation time $\lambda$.}
\label{fig:figS1}
\end{figure*}

Here, we characterize the rheology of our polymeric solutions by measuring the viscosity and relaxation time.
The viscosity $\eta$ is measured as a function of the imposed shear rate $\dot{\gamma}$ using a rheometer (Anton Paar, MCR 502 with CP50-1$^{\circ}$
cone-plate geometry).
For water, the viscosity remains constant around $\eta = 1$ mPa$\cdot$s (Fig.\,\ref{fig:figS1}a).
In contrast, the polymer solutions adopt a shear thinning behavior.
The viscosity remains constant, but starts to decrease beyond a certain shear rate. 
For the relaxation time, we measure the minimum thickness $h$ formed during the breakup of drop from a needle with diameter $d$ (Fig.\,\ref{fig:figS1}b).
For water, the breakup occurs within milliseconds and the thickness is characterized by a power-law.
Conversely, the breakup of polymer solution is largely delayed due to the axial stretching of the polymers.
Consequently, the an almost uniform thread is formed during breakup, whose thickness $h$ decreases with time.
The thread thickness for polymeric solutions follows an exponential decay $h/d \sim \exp(-t/3\lambda)$. 
We thus measure the relaxation by fitting exponential decay curves to the thread thickness.

\section{Oldroyd-B model}
\label{app:Oldroyd}
In the elastic limit of the Oldroyd-B model and using $ \mathbf{u} = r_0 \mathbf{e}_r$, the azimuthal component of the conformation tensor $A_{\theta\theta}$ is given by


 \begin{equation}
    \label{eq:Oldroyd_B_azimuthal}
    \frac{\partial A_{\theta\theta}}{\partial t} + u \frac{\partial A_{\theta\theta}}{\partial r}-\frac{2u}{r}A_{\theta\theta}  = 0,
\end{equation}
which corresponds to Eq.\eqref{eq:Oldroyd_B_azimuthalbis} of the main text.
In this limit, the polymers behave purely as elastic solids, allowing us to directly relate $A_{\theta\theta}$ to the elastic stretch experienced by the polymers~\cite{snoeijer2020relationship}.
The elastic stretch can be computed by describing the deformation with respect to an undeformed reference state, at which the elastic energy is at a minimum.
Introducing, the Lagrangian coordinates $\mathbf{R}$ in the undeformed reference state and the Eulerian coordinates $\mathbf{r}$ in the deformed state, the stretching of the polymers is given by the deformation gradient tensor $\mathbf{F} = \frac{\partial \mathbf{r}}{\partial \mathbf{R}}$.
For the azimuthal stretching considered here, the diagonal components of $\mathbf{F}$ correspond to the stretches along the principal directions.

In the absence of relaxation or initial pre-stretching, the conformation and deformation gradient tensors are related via  $\mathbf{A} = \mathbf{F}\cdot\mathbf{F}^{\mathrm{T}}$~\cite{eggers2020self,snoeijer2020relationship}.
Therefore, the conformation tensor becomes the square of the stretching, such that $A_{\theta\theta} = F_{\theta\theta}^2 =(r/R)^2$, where $r$ and $R$ are the radial Eulerian and Lagrangian coordinates respectively.
The mapping for liquid retraction during bubble coalescence can be solved exactly using volume conservation.
The total volume of the retracting liquid up to a position $r$ in both the reference and current state balance as
\begin{equation}
    \int_0^R H(\tilde{R}) \tilde{R}\,\mathrm{d}\tilde{R} = \int_{r_0(t)}^r h(\tilde{r},t) \tilde{r}\,\mathrm{d}\tilde{r}.
    \label{eq:mapping}
\end{equation}
In the reference state, the two spherical bubbles are in contact and the film thickness can be approximated with a parabola as $H(R) = a - \sqrt{a^2-R^2} \approx R^2/(2a)$.
Injecting this profile into Eq.~\eqref{eq:mapping}, the mapping between the Lagrangian and Eulerian coordinates follows $R = \left(8a \int_{r_0(t)}^{r} h(\tilde{r},t)\, \tilde{r}\mathrm{d}\tilde{r} \right)^{1/4}$.
As a result, the azimuthal component $A_{\theta\theta}$ is given by
\begin{equation}
    A_{\theta\theta} = \frac{r^2}{\left[8a \int_{r_0(t)}^{r} h(\tilde{r},t)\, \tilde{r}\mathrm{d}\tilde{r} \right]^{1/2}},
    \label{eq:A_theta_app}    
\end{equation}
which corresponds to Eq.~\eqref{eq:A_theta} of the main text.
One can verify that Eq.~\eqref{eq:A_theta_app} is an exact solution to Eq.~\eqref{eq:Oldroyd_B_azimuthal} where the liquid mass conservation equation $r\dot{h} + \frac{\partial}{\partial r}(r h u) = 0$ needs to be invoked.

\providecommand{\noopsort}[1]{}\providecommand{\singleletter}[1]{#1}%

\end{document}